\def\year{2019}\relax
\documentclass[letterpaper]{article}
\usepackage{aaai19}
\usepackage{times}
\usepackage{helvet}
\usepackage{courier}
\usepackage{url}
\usepackage{graphicx}
\usepackage{url}
\usepackage{multicol}
\usepackage{tabularx}
\usepackage{xcolor}
\usepackage{times}
\usepackage{helvet}
\usepackage{courier}
\usepackage{float}
\usepackage{multirow}
\usepackage{flushend}
\title{HoaxItaly: a collection of Italian disinformation \\ and fact-checking stories shared on Twitter in 2019}
\author{
Francesco Pierri,\textsuperscript{\rm 1}
Alessandro Artoni,\textsuperscript{\rm 1}
Stefano Ceri,\textsuperscript{\rm 1}\\
\textsuperscript{\rm 1}Dipartimento di Elettronica, Informazione e Bioingegneria\\
Politecnico di Milano\\
Via Giuseppe Ponzio, 34, I-20133 Milano, Italy\\
\{firstname.lastname\}@polimi.it
}
\begin{document}

\maketitle
\begin{abstract}
We released over 1 million tweets shared during 2019 and containing links to thousands of news articles published on two classes of Italian outlets: (1) disinformation websites, i.e. outlets which have been repeatedly flagged by journalists and fact-checkers for producing low-credibility content such as false news, hoaxes, click-bait, misleading and hyper-partisan stories; (2) fact-checking websites which notably debunk and verify online news and claims. The dataset, which includes also title and body for approximately 37k news articles, is publicly available at \url{https://doi.org/10.7910/DVN/PGVDHX}.
\end{abstract}

\section{Introduction}
We are witnessing an ever growing concern over the presence and the influence of deceptive and malicious information spreading on online social media \cite{Pierri2019}\cite{allcott2017}. Researchers use different terms to indicate the same issue, namely disinformation, misinformation, propaganda, junk news and click-bait. In this work we use the word \textit{disinformation}, rather than the more popular "fake news", to refer to a variety of low-credibility content which comprises false news intended to harm, misleading and non-factual reporting, hyper-partisan news and propaganda, and unverified rumors \cite{PierriScirep2019}.

Most of the research on this issue has focused on past 2016 US Presidential elections: it has been shown that false news spread deeper and faster than reliable news, with social bots and echo chambers playing a primary role in the diffusion \cite{Vosoughi18}\cite{Vosoughi18}\cite{botnature}\cite{misinformation}; however, misleading information usually amounts to a negligible fraction of overall online news, and it is mostly shared by old and conservative leaning people, highly engaged with politics \cite{Grinberg}\cite{Bovet2019}.

Nevertheless, disinformation spreading on social platforms has been also reported in European countries in different circumstances, including 2016 Brexit \cite{bastos2019}, 2017 French Presidential Elections \cite{jnFRA}\cite{ferrara2017}, 2017 Catalan referendum \cite{spagna}, 2018 Italian General elections \cite{doesfakenews} \cite{giglietto2018} and 2019 European elections \cite{jnSWE}\cite{jnGE}\cite{jnFRA}\cite{jnEU2019}\cite{PierriArtoni2019}. 

For what concerns Italy, where according to Reuters trust in news is today particularly low \cite{reuters2019}, previous research has highlighted the existence of segregated communities \cite{quattrociocchi2017} when it comes to consume online news, and remarkable exposure to online disinformation was discovered in the run-up to 2018 Italian General elections \cite{giglietto2018}, a result which was also substantiated in a report of the Italian Authority for Communications Guarantees (AGCOM) \cite{agcom}. Besides, Facebook has recently shut down several pages and accounts for violating platform's terms of use, thanks to the insights provided by Avaaz \cite{avaaz}, which revealed the existence of a network of agents spreading low-credibility and inflammatory content about controversial themes such as immigration, national safety and anti-establishment.

In the following we first describe previous research contributions that relate to our work; next we provide a description of the collection procedure and the dataset itself, which is available at \url{https://doi.org/10.7910/DVN/PGVDHX}; finally we briefly describe two applications of this dataset and we conclude.

\section{Related work}
There are a few references in the literature which relate to this work.

As the title of this paper suggests, we based our contribution on \textbf{Hoaxy} platform \cite{hoaxy}\cite{hoaxy2}, which has been tracking--since 2016--the diffusion on Twitter of hundreds of thousands of news articles published on US disinformation and fact-checking websites, collecting millions of tweets containing explicit links to these domains. Authors provide an API to easily download their data (available at \url{https://rapidapi.com/truthy/api/hoaxy}), and also the source code for developing the collection pipeline (available at \url{https://github.com/IUNetSci/hoaxy-backend}), which we implemented starting from our own list of Italian sources.

\textbf{NELA-GT-2018} \cite{nela} is a large political news data set which contains 700k articles collected in 2018 from almost 200 US news outlets which include mainstream, hyper-partisan, and conspiracy sources. Authors used 8 different assessment sites to assess veracity, reliability, bias, transparency, adherence to journalistic standards, and consumer trust of these sources. %The dataset can be found at \url{https://doi.org/10.7910/DVN/ULHLCB}.

Authors in \cite{brena} built a pipeline for collecting data describing news sharing behaviour on Twitter: they take as input a list of news sources and collect articles shared on Twitter, as well as metadata about users sharing such stories. They released an associated dataset concerning the US landscape which contains approx. 1 million tweets and 300k articles.%, and which is available at \url{https://doi.org/10.7910/DVN/5XRZLH}.

Authors in \cite{gullible} provide a dataset of thousands of so-called \textit{credulous} Twitter users, i.e. human-operated accounts who exhibit a high percentage of bots among their followings. These might be particularly exposed to the harmful activities of social bots and they could become spreaders of misleading information. Authors also provide a supervised classifier to accurately recognize such users with accuracy up to 93.27\%. %The dataset is available at \url{https://toffee.imtlucca.it/datasets}.

\begin{figure}[!t]
    \centering
    \includegraphics[width=\linewidth]{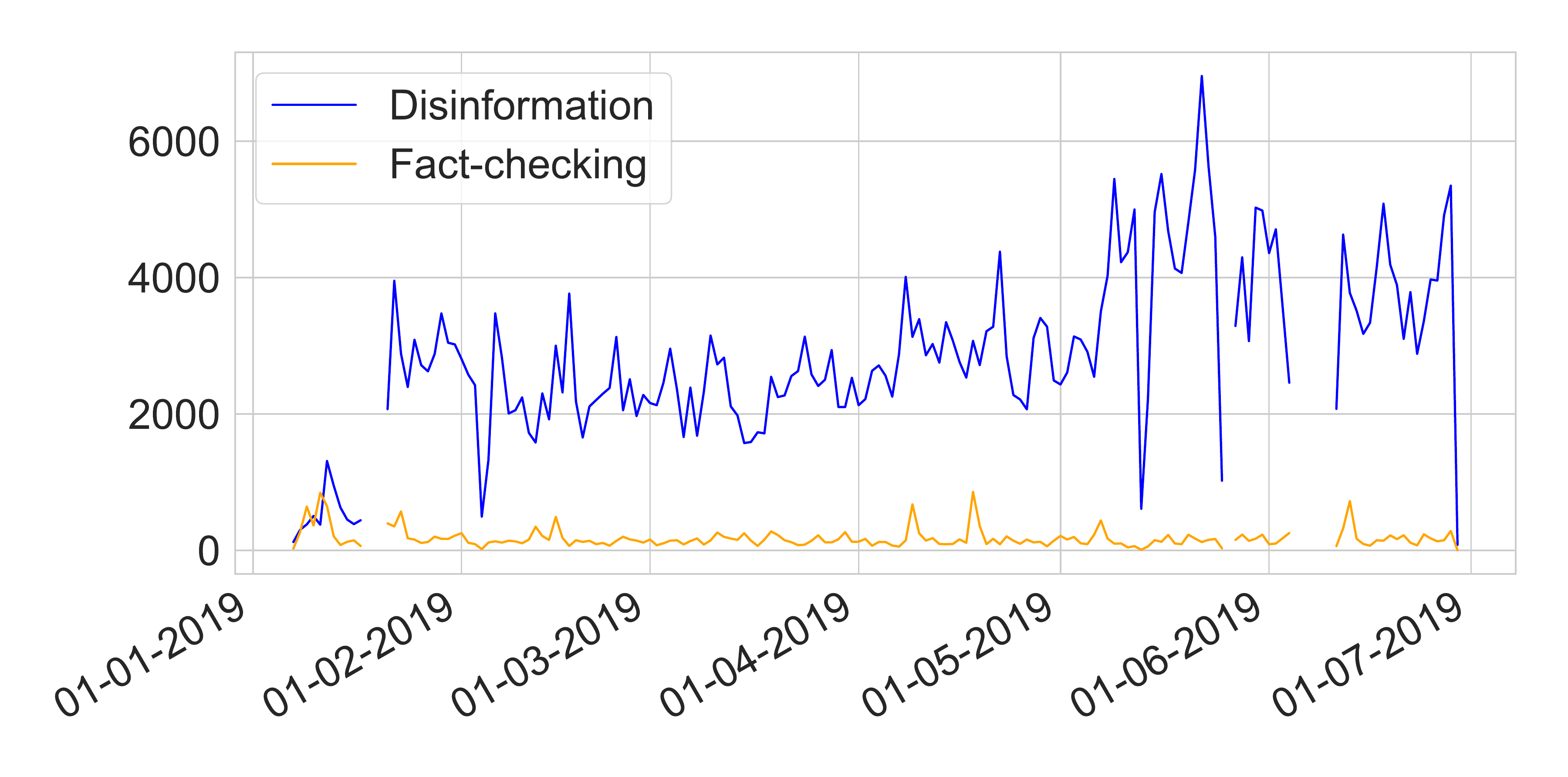}
    \\
    a) From January to July 2019
    \\
    \includegraphics[width=\linewidth]{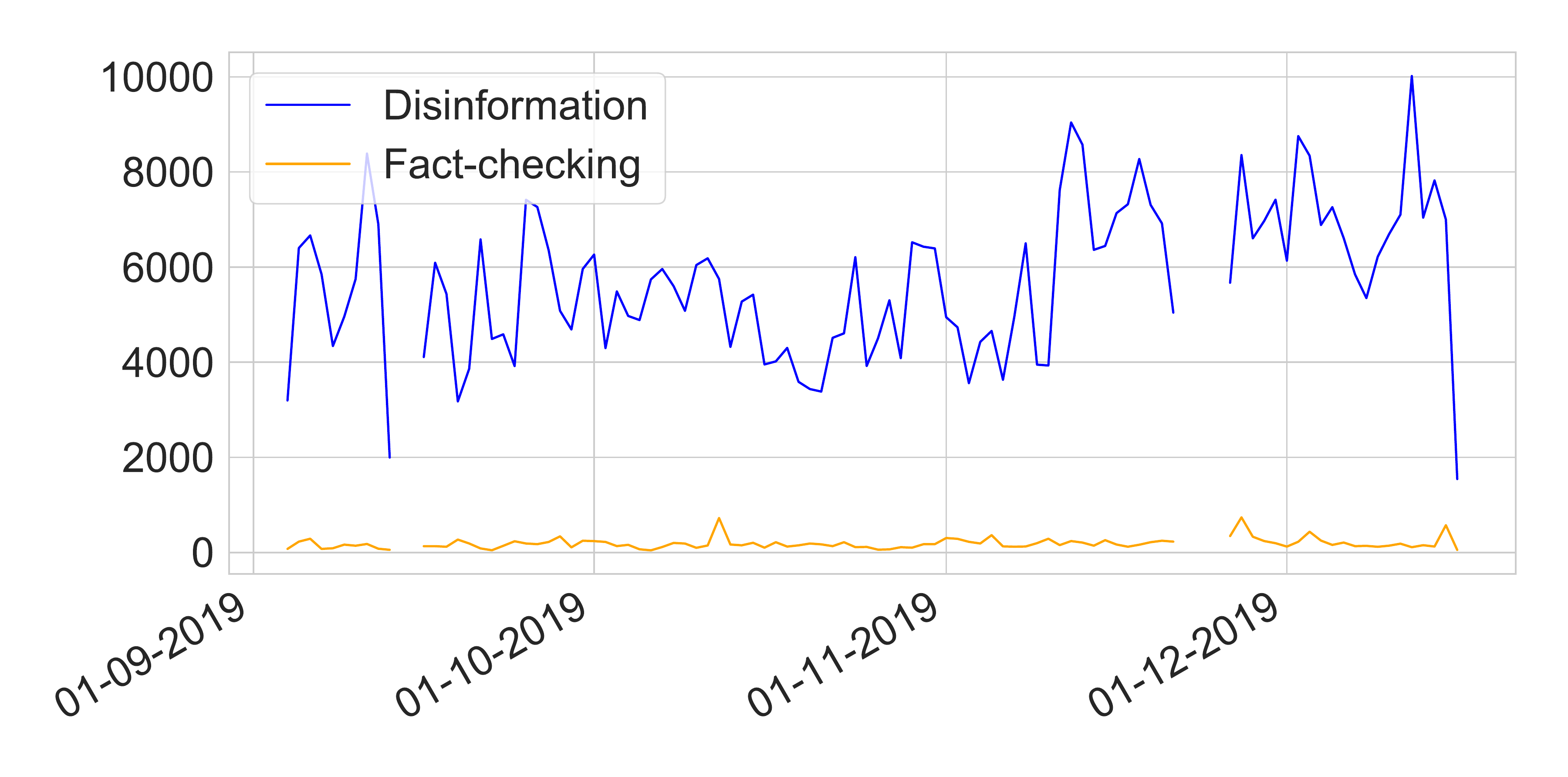}
    \\
    b) From September to December 2019
    \caption{Time series (in two halves) of the number of collected tweets, on a daily basis, for disinformation (blue) and fact-checking articles (orange). }
\end{figure}
\section{Data collection}
As previously mentioned, we referred to Hoaxy \cite{hoaxy} to setup our collection pipeline. We used Twitter Streaming API (via \verb|tweepy| Python package) from January to December 2019 to filter tweets containing URLs of 66 Italian disinformation outlets\footnote{A complete list of  these websites is available at \url{https://docs.google.com/spreadsheets/d/1YRNFXrr47w8pa1j3GmPnBjtIarf9_xFAHPT0GN1_lDQ}.} and a few fact-checking agencies, namely PagellaPolitica\footnote{www.pagellapolitica.it}, Bufale.net\footnote{www.bufale.net}, Butac.it\footnote{www.butac.it} and Lavoce.info\footnote{www.lavoce.info}. We referred to the blacklists compiled by these websites in order to build our list of untrustworthy websites, i.e. outlets who notably publish hoaxes and false news, but also conspiracy theories, hyper-partisan news, clickbait and pseudoscience.

We used web domains as \verb|track| parameter as described in Twitter documentation\footnote{\url{https://developer.twitter.com/en/docs}} to capture all related tweets, e.g. "voxnews info OR ilprimatonazionale it OR ...". We also built our own crawler in Python to scrape news articles and perform URL canonicalization. 

We first remark that we can only release tweet IDs--which can be used to retrieve original tweets via Twitter API using the \verb|status/lookup| Twitter endpoint--in accordance with Twitter terms of use, whereas we provide title and body for news articles which are publicly accessible.
Secondly, we are aware of the limitation of Twitter Streaming endpoint, which cuts out matched tweets when they reach 1\% of daily volume of shared tweets \cite{twittersample}; however, we did not incur in missing data as we collected roughly $\cdot10^5$ tweets per day, which is far less than the 1\% of current daily volume of tweets\footnote{\url{https://www.internetlivestats.com/twitter-statistics/}} that exceeds $2\cdot10^8$ shared tweets.

\begin{table}[!t]
\centering
\resizebox{\linewidth}{!}{%
\begin{tabular}{l|lll} \textbf{Domain} &  \textbf{Articles} & \textbf{Tweets} & \textbf{Users} \\ \hline
Fact-checking & 3,566 & 50,662 & 17,155 \\ \hline
Disinformation & 32,686 & 1,068,102 & 45,112 \\ \hline
\end{tabular}%
}
\caption{Breakdown of the dataset in terms of articles, tweets and users.}
\end{table}

\begin{figure}[!t]
    \centering
    \includegraphics[width=0.9\linewidth]{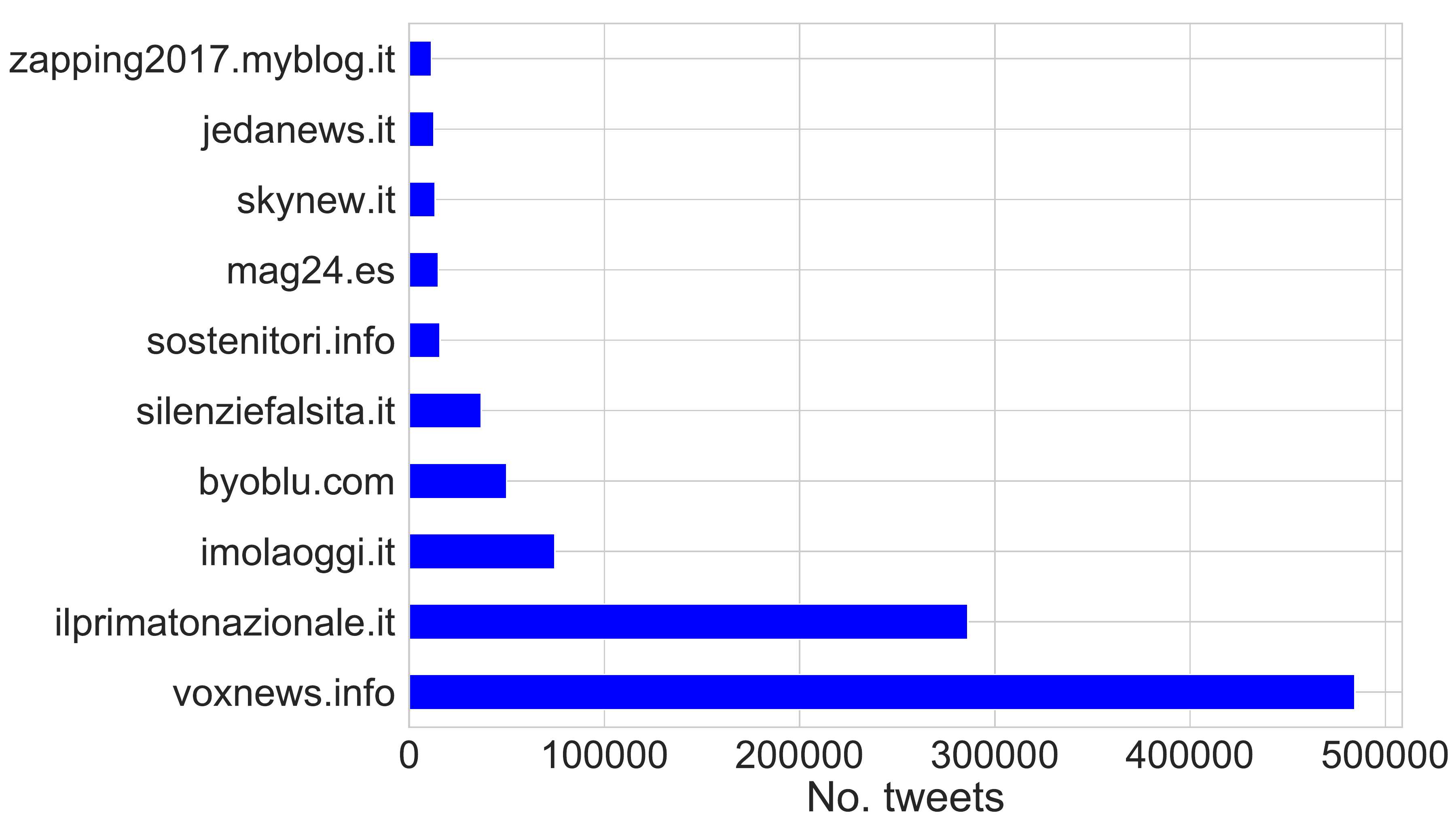}
    \\
    \footnotesize{\textbf{a)} Top-10 disinformation sources}
    \\
    \includegraphics[width=0.9\linewidth]{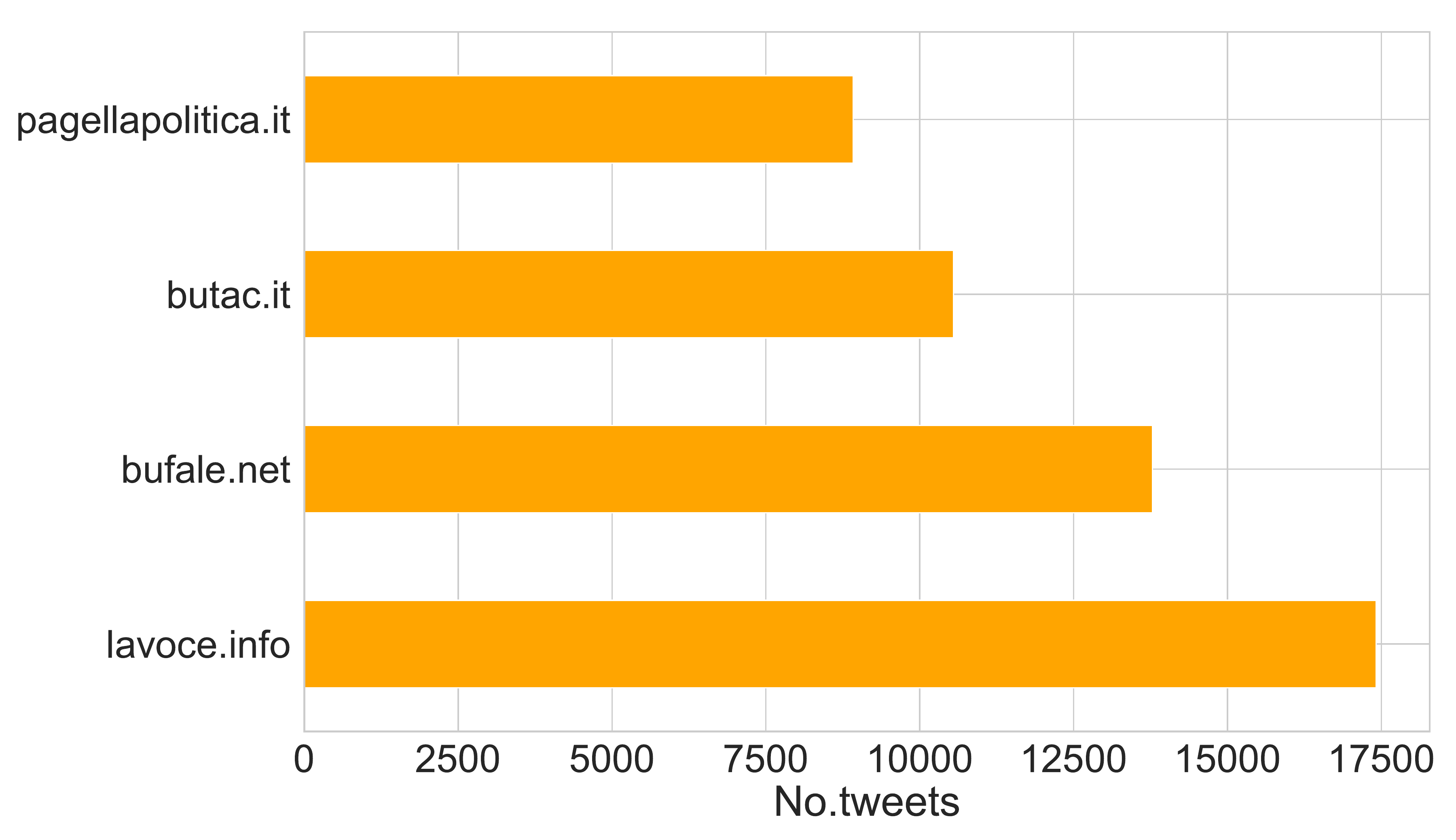}
    \\
    \footnotesize{\textbf{b)} fact-checking sources}
    \caption{Distribution of the number of articles for \textbf{a)} disinformation and \textbf{ b)} fact-checking sources.}
    \label{fig:breakdown-ita}
\end{figure}

\section{Data description}
In Figure 1 we show the time series of collected tweets on a daily basis, for both disinformation (black) and fact-checking news (orange). We collected the data in two phases: the first goes from January to mid-July, when we experienced a pause in the collection; we then resumed it at the beginning of September and ran it until December. Blank spaces correspond to temporary network failures. 
A breakdown of the dataset in terms of articles, tweets and users is provided in Table 1.
We can notice that the presence of fact-checking stories is negligible compared to disinformation news--less than 5\% of total tweets contain a link to fact-checking stories which amount to less than 10\% of total shared articles--in line with findings in the US \cite{hoaxy}.

% \begin{figure*}[!t]
%     \centering
%     \includegraphics[width=0.44\linewidth]{claim_fig2.pdf}
%     % \footnotesize{\textbf{a)} Number of tweets per user}
%     % \\
%     \includegraphics[width=0.44\linewidth]{fact_fig2.pdf}
%     % \\
%     % \footnotesize{\textbf{b)} Number of unique articles per user}
%     \caption{Title for Top-10 shared articles in terms of tweet for \textbf{(left)} disinformation and \textbf{(right)} fact-checking domains.}
% \end{figure*}

In Figure 2 we show the distribution of the number of shared tweets for Top-10 disinformation sources (which account for over 99\% of related tweets) and fact-checking sources, whereas in Table 3 we show Top-10 disinformation and fact-checking articles w.r.t the total number of shares.

We can notice that there are only a handful of highly active outlets which are responsible for most of the misleading information spreading on Twitter. In particular, "voxnews.info" (and most other websites in the Top-10) notably publish hoaxes and fake news\footnote{See for example the list of debunked articles at \url{https://www.open.online/?s=voxnews}} which usually involve immigrants, refugees and national safety; "ilprimatonazionale.it" is the reference news outlet for Italian far-right community (and official website of former party Casa Pound), and it has often been compared to BreitbartNews for its propaganda and hyper-partisan reporting; finally, "byoblu.com" is rather involved with the Italian euro-skeptical community and it regularly spreads conspiracy theories.

\section{Applications}
In the following we describe two applications of our dataset referring to our previous works, namely (1) a detailed investigation of disinformation spreading on the Italian Twitterverse in the run-up to 2019 European Parliament elections \cite{PierriArtoni2019}, and (2) a classification of mainstream and disinformation news based on multilayer networks \cite{Pierri2020}.

\begin{figure}[!t]
    \centering
    \includegraphics[width=0.8\linewidth]{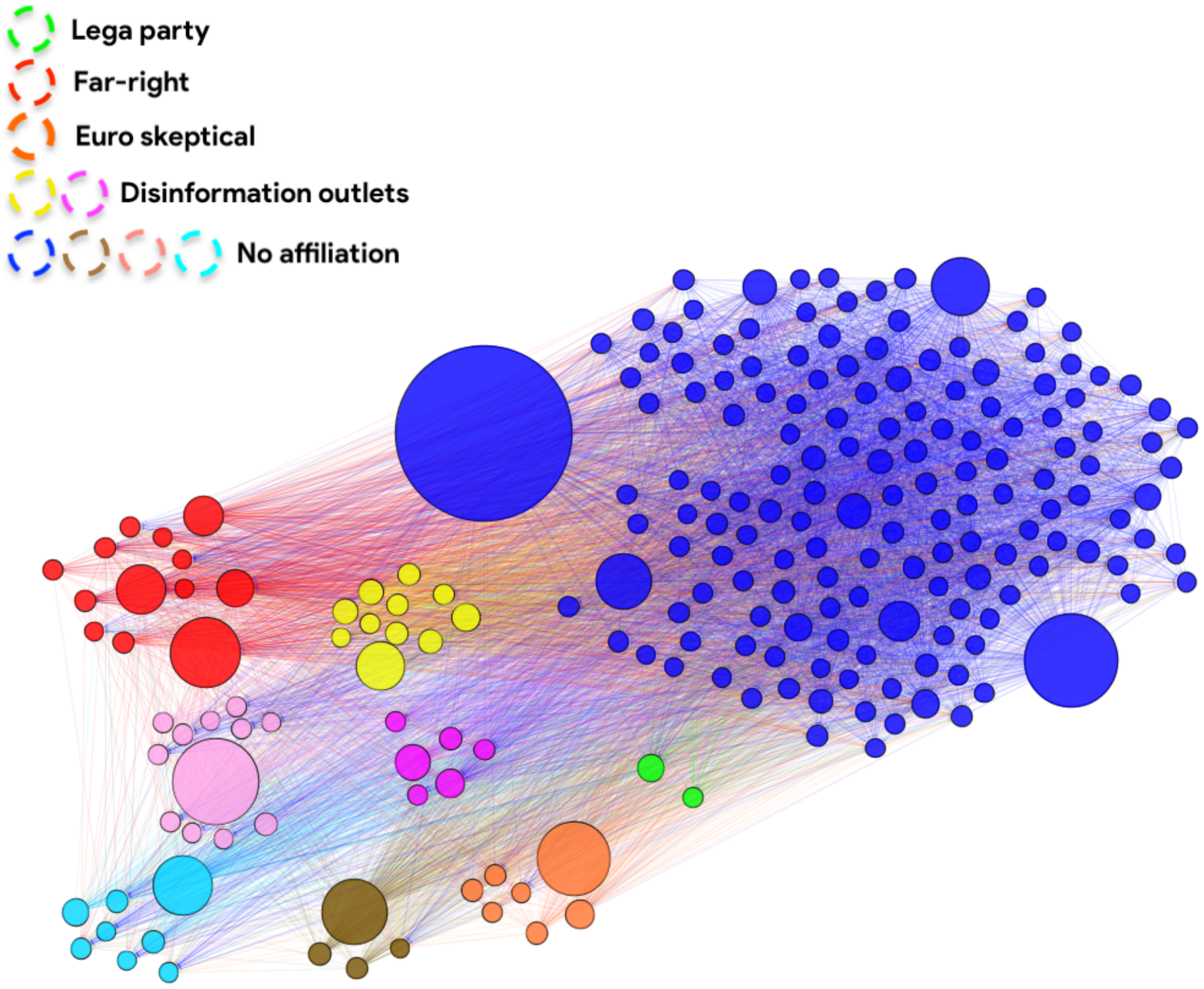}
    \\
    \footnotesize{\textbf{a)} Main K-core of the disinformation diffusion network}
    \\
    \includegraphics[width=\linewidth]{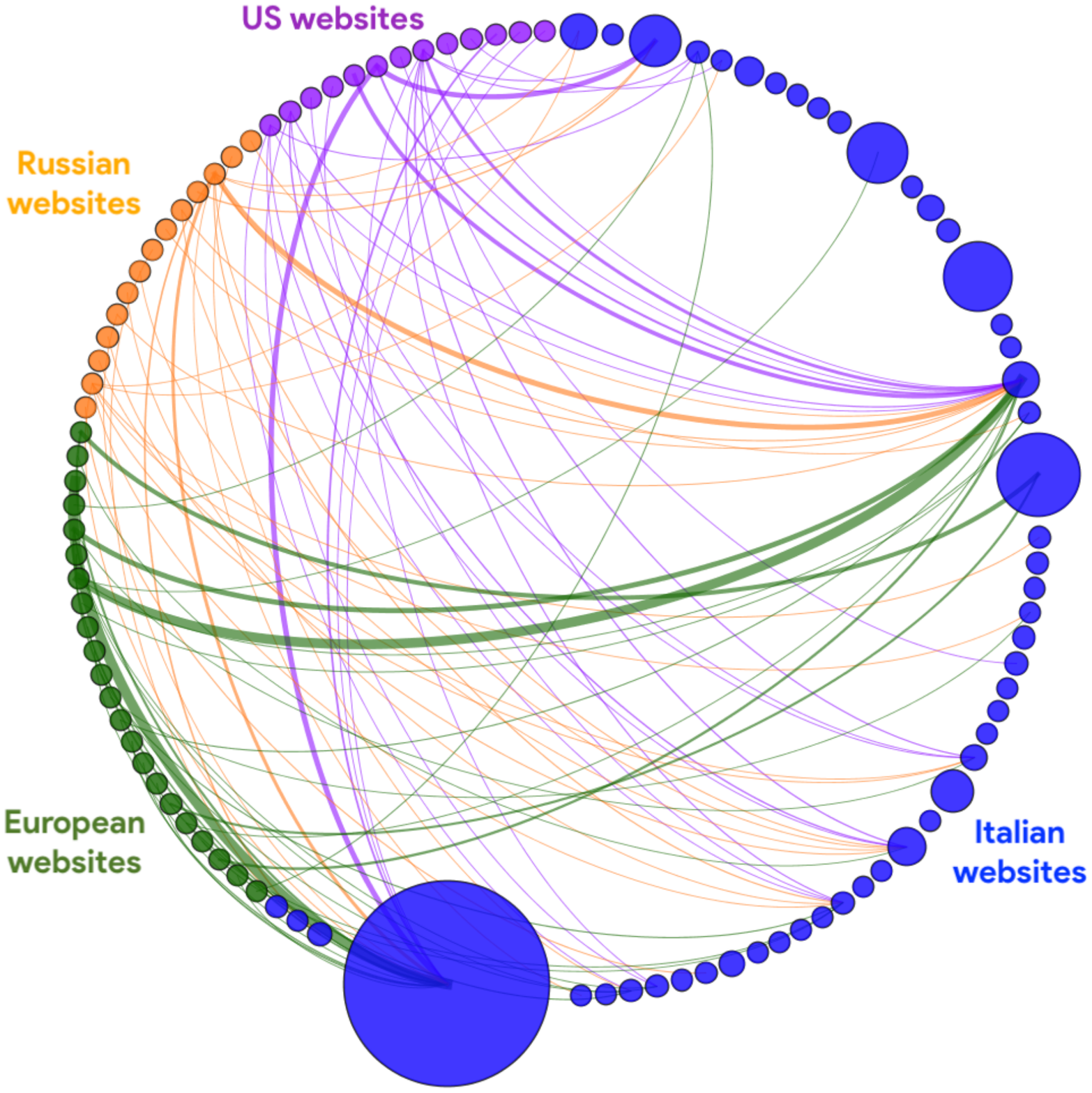}
    \\
    \footnotesize{\textbf{b)} Network of hyperlinks between disinformation websites from different countries}
    \caption{A visualization of \textbf{a)} the main K-core of the disinformation diffusion network and \textbf{b)} the network of hyperlinks between outlets from different countries. Colors indicate different groups according to a community detection algorithm in \textbf{a)}, and the country of provenance in \textbf{b)}.}
\end{figure}

\textbf{Disinformation in the run-up to 2019 European Parliament elections.} We analyzed the 5-month period preceding the 2019 European Parliament elections (which took place on the May, 26th) to investigate the presence and influence of disinformation in the Italian Twitterverse. The dataset analyzed corresponds to part of the collection we release in this paper.

We observed that most of the deceptive information was shared by a few outlets, and a dictionary-based topic analysis revealed that misleading news mostly focused on on controversial and polarizing topics of debate such as immigration, national safety and (Italian) nationalism. 

We also analyzed the main K-core \cite{k-core} of the diffusion network and noticed that the spread of disinformation on Twitter was confined in a limited community, strongly (and explicitly) related to the Italian conservative and far-right political environment; a network dismantling analysis showed that this could be easily dismantled with several strategies based on node centrality measures. 

Finally, we searched hyperlinks contained in news articles and discovered connections between different disinformation outlets across Europe, U.S. and Russia, which often featured similar, even translated, articles. 

We show in Figure 3 the core of the global diffusion network (where colors indicate groups identified with a community-detection algorithm) and the network of hyperlinks between websites (where colors indicate different countries). We refer the interested reader to \cite{PierriArtoni2019} for more details on the results.

\begin{figure}
    \centering
    \includegraphics[width=\linewidth]{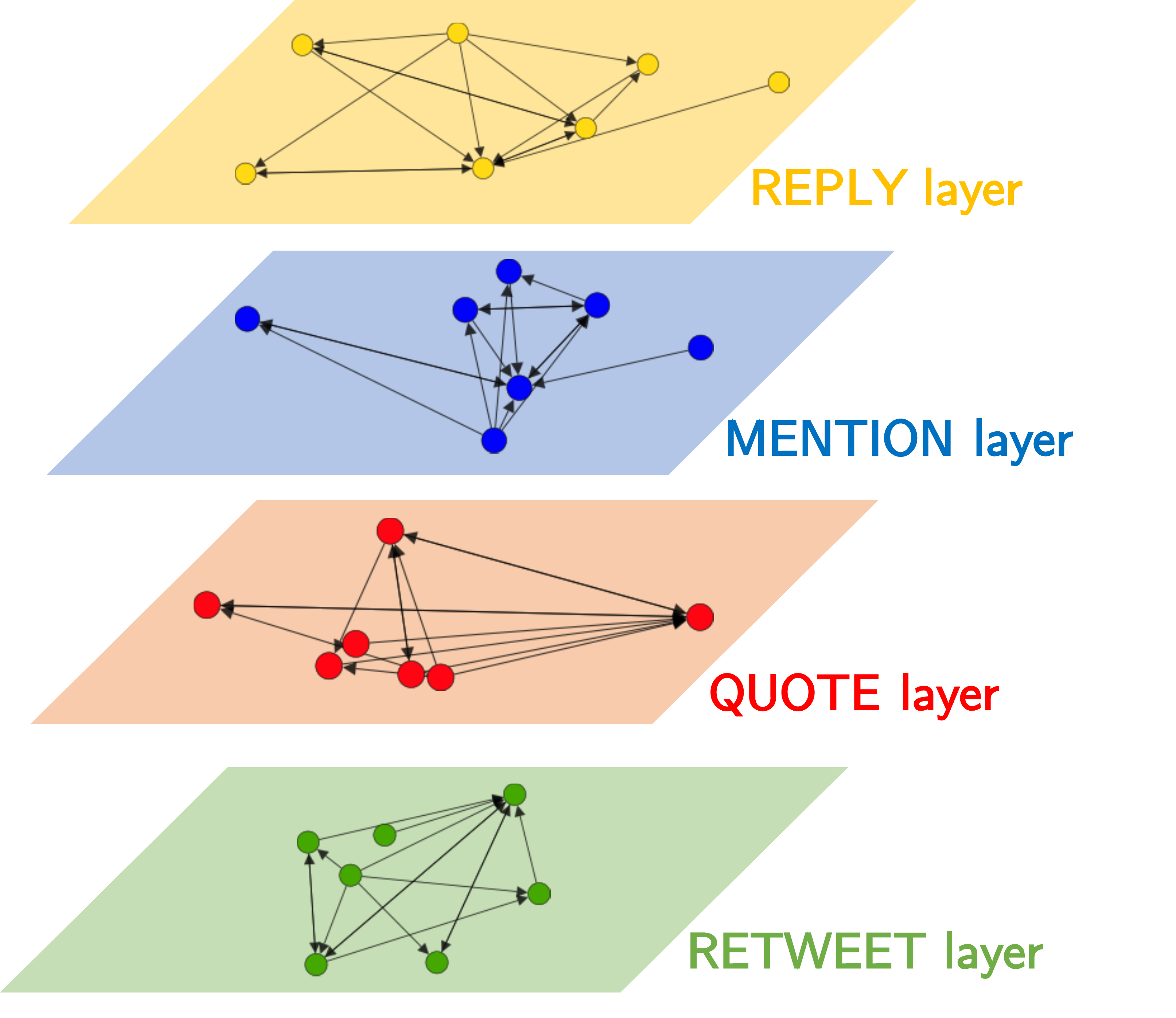}
    \caption{An example of multi-layer diffusion network on Twitter.}
\end{figure}

\textbf{Disinformation detection via multilayer diffusion networks.} We developed a machine learning framework to classify news articles published on \textit{disinformation} vs \textit{mainstream} news outlets based on Twitter diffusion networks. 

For each article we built a multi-layer diffusion network (see Figure 4) based on different social interactions on Twitter, namely tweets, retweets, mentions, replies and quotes. We selected a set of global network properties, which quantify different aspects of the sharing process, to encode different networks with a tuple of features. These range from traditional indicators, e.g. network density, number of strong/weakly connected components and diameter, to more elaborated ones such as main K-core number \cite{k-core} and structural virality \cite{goel}.

We performed classification experiments with a simple Logistic Regression model using two different datasets of mainstream and disinformation news shared on Twitter respectively in the United States and in Italy during 2019 (the latter is part of the collection we release in this work). 

We were able to classify credible vs non-credible articles with high accuracy (AUROC up to 94\%), and we observed that the layer of mentions alone conveys useful information for the classification, suggesting a different usage of mentions when sharing news belonging to the two news domains. We also highlighted that most discriminative features, which are relative to the breadth and depth of largest cascades in different layers, are the same across the two countries. We refer the interested reader to \cite{Pierri2020} for more details.

\section{Conclusions}
We released a large-scale dataset of tweets and articles concerning disinformation and fact-checking stories circulating on the Italian Twitterverse in 2019, and we provided a related descriptive statistics of sources, articles, tweets and users. We also described two past applications of this dataset which concern (1) the analysis of disinformation spreading in Italy before 2019 European Parliament elections and (2) an approach to detect disinformation based on Twitter diffusion networks represented in a multi-layer flavor.

We hope that the research community might successfully embody this data as to further investigate the issue of malicious information outside the United States context, where most of the research focus has been devoted in the past.

\begin{table*}[!t]
\centering
\resizebox{\linewidth}{!}{%
\begin{tabular}{lll} \textbf{Title} & \textbf{Website} & \textbf{No. tweets}\\ \hline
Chi sono davvero le sardine “apartitiche”. La schedatura dei leader nazionali e locali & ilprimatonazionale.it & 2490
\\ \hline
La Consulta boccia l’omogenitorialità: “Le coppie gay non sono famiglie” & ilprimatonazionale.it & 2200
\\ \hline
La figuraccia di Italia Viva: firme false nella petizione contro i “profili fake” & ilprimatonazionale.it & 2047
\\ \hline
CasaPound vince la causa contro Facebook. Il giudice ordina la riattivazione della pagina & ilprimatonazionale.it & 1978
\\ \hline
Più della metà degli account Twitter che seguono Boldrini e Saviano sono falsi & ilprimatonazionale.it & 1928
\\ \hline
Giulia Berberi, medico della nave Alex? E’ una dentista & ilprimatonazionale.it & 1891
\\ \hline
Matteo Salvini, frena tutti: “Non cambio idea sui migranti, i porti restano chiusi” & lettoquotidiano.it & 1846
\\ \hline
Pakistano violentatore, la giudice pro Ong lo aveva protetto perché lui dichiarava di essere gay & ilprimatonazionale.it & 1690
\\ \hline
La Luiss fa fuori il prof Gervasoni: “Io cacciato per un tweet “sovranista”” & ilprimatonazionale.it & 1553
\\ \hline
11 barconi tunisini assaltano Lampedusa, poliziotto: “Sbarcano galeotti tunisini e TG ve lo nascondono” & voxnews.info & 1477
\\ \hline
\end{tabular}%
}
\caption{Top-10 disinformation articles per number of shared tweets.}
\end{table*}

\begin{table*}[!t]
\centering
\resizebox{\linewidth}{!}{%
\begin{tabular}{lll} \textbf{Title} & \textbf{Website} & \textbf{No. tweets}\\ \hline
Fact-checking: i cartelli di Salvini sull’immigrazione a Porta a Porta & pagellapolitica.it & 1395
\\ \hline
No, Laura Boldrini non ha occupato il posto dei disabili in aereo & pagellapolitica.it & 898
\\ \hline
Le spose bambine Vol.2 & butac.it & 797
\\ \hline
Salvini diffonde una bufala sull’agente preso a morsi da un detenuto: lo conferma la sorella & bufale.net & 712
\\ \hline
Quanto costa allo Stato un cervello in fuga? & pagellapolitica.it & 534
\\ \hline
NOTIZIA VERA Quando Salvini disse a Iliaria Cucchi “Mi fa schifo”. Ecco l’audio & bufale.net & 471
\\ \hline
Se Di Battista svela il complotto sul franco Cfa & lavoce.info & 469
\\ \hline
Il 66\% delle bufale politiche arrivano dai fan di Salvini e Meloni: i dettagli sul nuovo studio & bufale.net & 469
\\ \hline
L’effetto Salvini sugli sbarchi conta metà dell’effetto Minniti & lavoce.info & 423
\\ \hline
Di Maio confuso sui soldi alle banche & lavoce.info & 378
\\ \hline
\end{tabular}%
}
\caption{Top-10 fact-checking articles per number of shared tweets.}
\end{table*}

\section{Acknowledgements}
F.P. and S.C. are supported by the PRIN grant HOPE (FP6, Italian Ministry of Education). S.C. is partially supported by ERC Advanced Grant 693174.
Authors are grateful to Hoaxy team of developers at Indiana University and to PagellaPolitica.it for their support.

\bibliographystyle{aaai}
\bibliography{bib.bib}

\end{document}